\documentclass[aps,prb,twocolumn,superscriptaddress,groupedaddress]{revtex4}  

\usepackage{graphicx}  
\usepackage{dcolumn}   
\usepackage{bm}        
\usepackage{amssymb}   
\usepackage{amsmath}
\usepackage{braket}
\usepackage{fixltx2e} 

\usepackage{xcolor} 

\newcommand{\biblio}{/Users/marco/Desktop/GlobalBiblio}

\mathchardef\mhyphen="2D

\begin{document}

\title{Impact of valley phase and splitting on readout of silicon spin qubits}
\author{M.L.V.Tagliaferri}
\email{m.l.v.tagliaferri@tudelft.nl}
\affiliation{QuTech, TU Delft and Kavli Institute of Nanoscience, PO Box 5046, 2600GA, Delft, The Netherlands}
\author{P.L. Bavdaz}
\affiliation{QuTech, TU Delft and Kavli Institute of Nanoscience, PO Box 5046, 2600GA, Delft, The Netherlands}
\author{W. Huang}
\affiliation{Centre for Quantum Computation and Communication Technology, School of Electrical Engineering and Telecommunications, The University of New South Wales, Sydney 2052, Australia}
\author{A.S. Dzurak}
\affiliation{Centre for Quantum Computation and Communication Technology, School of Electrical Engineering and Telecommunications, The University of New South Wales, Sydney 2052, Australia}
\author{D. Culcer}
\affiliation{School of Physics, The University of New South Wales, Sydney 2052, Australia}
\affiliation{ARC Centre of Excellence in Future Low-Energy Electronics Technologies, Sydney 2052, Australia}
\author{M. Veldhorst}
\email{m.veldhorst@tudelft.nl}
\affiliation{QuTech, TU Delft and Kavli Institute of Nanoscience, PO Box 5046, 2600GA, Delft, The Netherlands}

\begin{abstract}
We investigate the effect of the valley degree of freedom on Pauli-spin blockade readout of spin qubits in silicon. The valley splitting energy sets the singlet-triplet splitting and thereby constrains the detuning range. The valley phase difference controls the relative strength of the $\textit{intra-}$ and $\textit{inter}$-valley tunnel couplings, which, in the proposed Pauli-spin blockade readout scheme, couple singlets and polarized triplets, respectively. We find that high-fidelity readout is possible for a wide range of phase differences, while taking into account experimentally observed valley splittings and tunnel couplings. We also show that the control of the valley splitting together with the optimization of the readout detuning can compensate the effect of the valley phase difference. To increase the measurement fidelity and extend the relaxation time we propose a latching protocol that requires a triple quantum dot and exploits weak long-range tunnel coupling. These opportunities are promising for scaling spin qubit systems and improving qubit readout fidelity.
\end{abstract}

\maketitle

\section{Introduction}
The experimental demonstration of high-fidelity quantum dot qubits with long-coherence \cite{veldhorst_addressable, tarucha2017_F99p9} that can be coupled to perform two-qubit logic gates \cite{veldhorst2015twoqubit,petta_CNOT,watson2017programmable} and used to execute small quantum algorithms\cite{watson2017programmable} has positioned silicon as a promising platform for large-scale quantum computation. Building upon these advances, exciting new directions forward have been proposed\cite{Loss1998DiVincenzo,Kane,taylor2005fault,petersson2012circuit,hill2015architecture,pica2016surface,tosi_processor,baart2017coherent}, that exploit uniformity \cite{Roy_architecture}, robustness against thermal noise \cite{Lieven_interfacing}, or semiconductor manufacturing \cite{veldhorst_CMOS}, and aim for   operation of quantum error correction codes\cite{Helsen_QEC} on qubit arrays.\\
Despite its promises, silicon poses specific challenges due to the six-fold degeneracy of its conduction band minimum in bulk. This degeneracy is lifted close to an interface, and a gap opens between perpendicular and in-plane valley doublets. Interfaces and gate electric fields cause coupling either in the same or different orbital levels. The same-doublet same-orbital coupling is the so-called valley mixing, while the other couplings are generally referred to as valley-orbit coupling\cite{zwanenburg2013silicon}. While silicon quantum dots can often be operated in vanishingly small valley-orbit coupling regime\cite{Morello_spinValley,friesen2010VOC,gamble2013disorder}, valley mixing can not be neglected. 
As a complex quantity it is determined by a phase, $\textit{valley phase}$, and a modulus, $\textit{valley splitting}$\cite{condition}. Typical valley splittings range from tens of neV to about 1 meV\cite{jiang2017valleyqubit,lim2011spin,yang2012orbital,veldhorst_addressable,kawakami2014qubit} and introduce new challenges for spin qubits defined in silicon quantum dots. The consequences of valley phase have been studied only in limited research, but found to be significant in valley-qubits\cite{wu2012coherent} and donors close to an interface\cite{calderon2008QDdonormodel}, while they strongly influence the exchange interaction\cite{zimmerman2017valley}. A crucial question is therefore how the valley physics impacts quantum computation with spins in silicon quantum dots.\\
Here, we investigate the effect of valley mixing on the dynamics between spins in silicon quantum dots. In particular, we study readout, now one of the most challenging operations for spin qubits. We concentrate on Pauli-spin blockade readout and show that high spin-to-charge conversion fidelity is achievable in a wide parameter range. This readout technique is considered in large-scale quantum computation proposals\cite{Lieven_interfacing, veldhorst_CMOS, Roy_architecture} since it requires few electron reservoirs and is compatible with moderate magnetic fields \cite{petta2005coherent, maune2012HRL}. However, in standard Pauli-spin blockade schemes the readout time is limited due to spin-relaxation \cite{johnson2005relaxationPSB,petta2005pulsedrelaxation,HansonReviewSpin}. Moreover, usually two spin states are projected on charge states which differ only by the electric dipole, thus leading to a readout fidelity which can be much smaller than the conversion fidelity\cite{fogartyPSB,harvey2017latching}. A possible solution is to exploit latching mechanisms in the pulsing scheme, which locks the charge in a long-lived metastable state, such that the total amount of electrons differs between the two final states. However, proposed schemes require an external reservoir\cite{studenikinEnhanced,china2017readout,tarucha2017readout,fogartyPSB,harvey2017latching}. Here we overcome these limitations and propose a protocol based on a triple quantum dot that enables to measure the charge states resulting from the Pauli-spin blockade spin-to-charge conversion with high fidelity.\\
This work is organized as follows. In Section \ref{sec:model}, we introduce the model describing a multi-valley two-electron double quantum dot and discuss Pauli-spin blockade readout. In Section \ref{sec:results}, we investigate how the valley phase difference and splitting energy impact the spin-to-charge conversion fidelity. We identify the conditions that enable readout fidelities beyond 99.9\%. A triple quantum dot scheme combining Pauli-spin blockade with long lived charge states is proposed and studied in Section \ref{sec:TQD}. We discuss the conclusions and opportunities in Section \ref{sec:conclusions}.

\section{Model}
\label{sec:model}
\subsection{Silicon Double Quantum Dot Hamiltonian}
\begin{figure}
\includegraphics[width=\columnwidth]{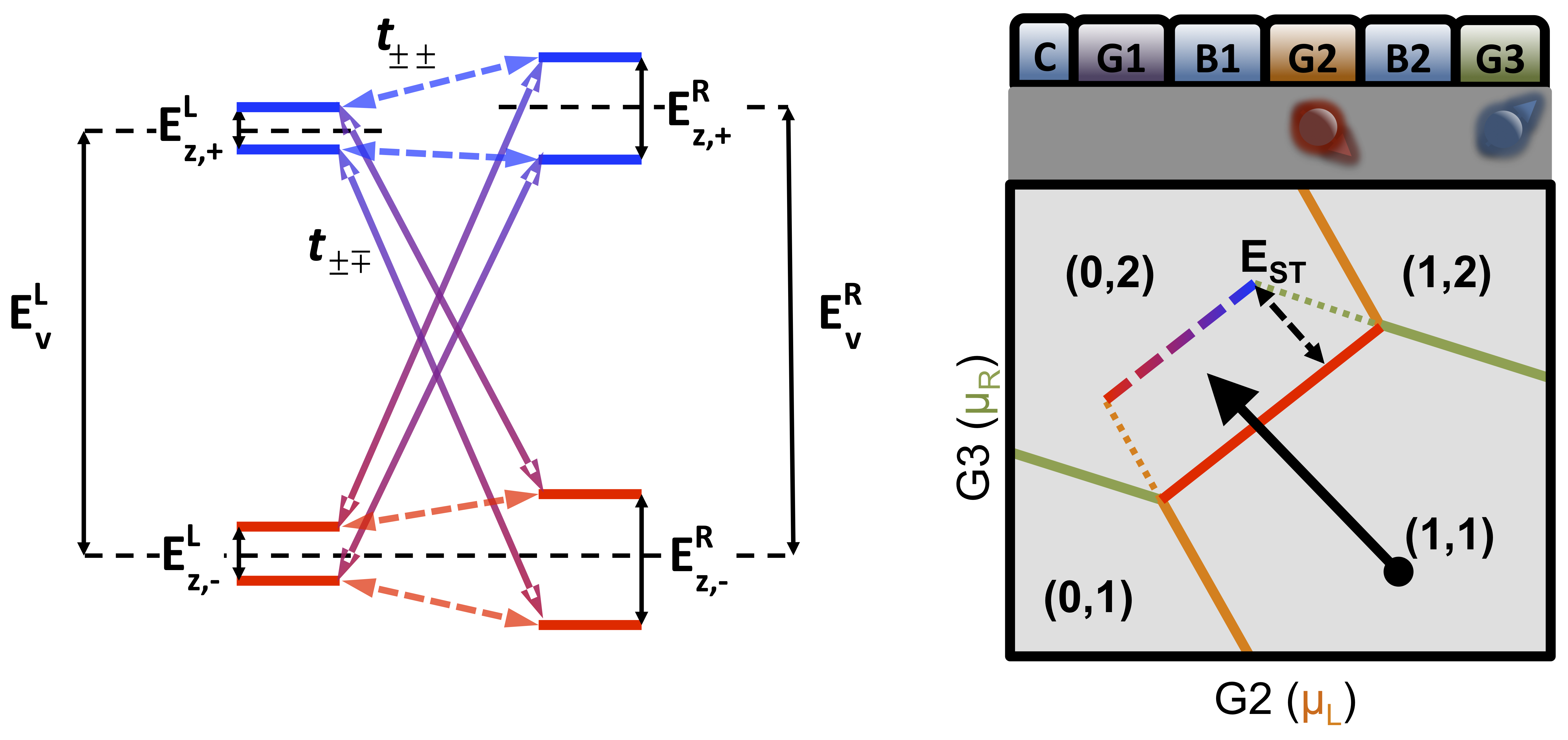}
\caption{a) Spin-valley single-particle energy levels of a silicon double quantum dot. The valley ground states are shown in red, while the excited valley states, separated by the dot-dependent valley splitting $E_v^{L(R)}$, are in blue. A dot and valley-dependent Zeeman energy (e.g. $E_{Z,-}^L$) splits the spin states. The constant color arrows represent the intravalley tunnel coupling $t_{\pm\pm}$, while intervalley coupling $t_{\pm\mp}$ arrows have a color gradient. b) Top: Cross-section of a schematic device. The confinement gate (C) defines the dots, plunger gates (G) accumulate the electrons and control the out-of-plane electric field, while the barrier gates (B) tune the tunnel coupling $t$. Bottom: Schematic stability diagram of a double quantum dot. Green and orange lines mark the dot-lead transitions. Interdot intervalley tunneling occurs along the dashed line, separated by the singlet-triplet energy $E_{ST}$ from the ground state line.}
\label{fig:model}
\end{figure}

The model developed in this section can be generalized to other doubly occupied multi-valley quantum dots, but here we restrict the discussion to quantum dots at the Si/SiO$_2$ interface. We consider a double quantum dot system as shown in Fig. \ref{fig:model}a and define the left quantum dot as target qubit and the right dot as ancilla qubit.\\
To describe the double quantum dot we consider ten single-particle spin-valley states: the four lowest orbital spin-valley states of each dot and the two lowest valley states in the first excited orbital of the ancilla qubit, that are needed to build the same-valley triplet states of the doubly occupied ancilla qubit. As shown in Fig. \ref{fig:model}b, the double dot is tuned by means of two plunger gates (G), controlling the energy levels, and two barrier gates (B), setting the interdot tunnel coupling. Furthermore we assume a valley splitting energy $E_v= [100\, \mu\rm{eV}, 1\, \rm{meV}]$ and an orbital splitting energy close to 10 meV, consistent with experimentally measured values\cite{lim2011spin,yang2012orbital,veldhorst_addressable}. The order of magnitude larger orbital splitting, together with operation at a small magnetic field, justifies the assumption of a negligibly small valley orbit coupling and pure valley mixing\cite{Morello_spinValley,zwanenburg2013silicon,friesen2010VOC}.\\
Each dot is described by the Hamiltonian $H_{0}^d=H_v^d+H_z^d+\delta_{d,R}H_o^R$, where $d=L,R$ is the dot label, $H_v^d$ describes the valley spectrum of the dot, $H_z^d$ the Zeeman splitting and $H_o^R$ the orbital levels of the right dot ($\delta_{d,R}$ is the Kronecker delta). In particular:

\begin{align}
H_v^d &= E_v^d\sum_{v=-,+}\delta_{v,+}\sum_{\substack{o=0,1\\ \sigma=\downarrow,\uparrow\\}}c_{\substack{d,o,\\ v,\sigma}}^\dagger c_{\substack{d,o,\\ v,\sigma}}
\;,
\label{eq:hamv}
\\
H_Z^d &= \frac{1}{2} \sum_{\sigma=\downarrow,\uparrow}(-1)^{\delta_{\sigma,\uparrow}}\sum_{\substack{o=0,1\\ v=-,+}}E_Z^{d,v}c_{\substack{d,o,\\ v,\sigma}}^\dagger c_{\substack{d,o,\\ v,\sigma}}
\;,
\label{eq:hamZ}
\\
H_o^R &= E_o^R\sum_{\substack{o=0,1}}\delta_{o,1} \sum_{\substack{v=-,+\\ \sigma=\downarrow,\uparrow}}c_{\substack{R,o,\\ v,\sigma}}^\dagger c_{\substack{R,o,\\ v,\sigma}}
\;,
\label{eq:hamdot}
\end{align}
where $o,v$ and $\sigma$ are the orbital, valley and spin labels respectively. The Zeeman splitting is defined as $E_Z^{d,v}=g_{d,v}\mu_BB_d$. In general the $g$-factor is valley and dot dependent due to spin-orbit coupling\cite{veldhorst2015gfactor,Carroll?}. Here we assume that the magnetic field is applied along one of the directions minimizing the spin-orbit coupling and assume a resulting vanishingly small strength\cite{ferdous2017interface,ferdous2017anysotropy,ruskov2017electron}. This assumption is further warranted by the possibility of low magnetic field operation when using Pauli-spin blockade readout (low field operations has several other advantages, see Refs. \citenum{Lieven_interfacing,Roy_architecture}). In this range, finite $\delta E^v_Z=E_Z^{R,v}-E_Z^{L,v}$ can be realized via nanomagnets, and we restrict to the case $E_Z^{R,v} > E_Z^{L,v}$. $H_v$ describes the mixing of the $k_{\pm z}$ bulk valleys due to the Si/SiO$_2$ interface and the electric field\cite{friesen2007valley,saraiva2011EMAtheory,culcer2010multivalley}. We consider dot-dependent valley splittings\cite{jiang2017valleyqubit} due to interface effects and local variations in electric field \cite{Morello_spinValley}. The valley coupling is $\Delta_v\equiv E_ve^{i\phi_D}$, whose modulus is the valley splitting energy and whose phase is the valley phase (i.e. the phase of the fast Bloch oscillations of the wave function)\cite{saraiva2009PRBR,friesen2006APLphase}. The valley eigenstates are of the form $D_{\pm}=(1/\sqrt{2})(D_z\pm e^{i\phi_D}D_{-z})$.\\
The two-electron double-dot Hamiltonian reads: 
\begin{equation}
H_{2e} = H_0+H_\epsilon+H_C+H_t.
\label{eq:2e}
\end{equation}
Here $H_0$ describes two non-interacting quantum dots. $H_\epsilon$ is the detuning term, describing the shift $\epsilon$ of the energy levels in the right dot with respect to the levels in the left dot, controlled by the gates. Referring to Fig. \ref{fig:model}b, this corresponds to increasing the voltage on G3. The third term $H_C$ accounts for the effect of the Coulomb potential $V_{ee}$. For the system considered here and within the Hund-Mulliken approximation, the Coulomb exchange integral $j=\bra{L_{z} R_{\pm z}}V_{ee}\ket{R_{z} L_{\pm z}}$ and the valley exchange integral $j_v=\bra{D_{\mp z} D_{\pm z}}V_{ee}\ket{D_{\pm z} D_{\mp z}}$ are negligible\cite{culcer2010roughness,Culcer_2electron,HadaEto-PRB}. Theoretical works have estimated $j\approx 1 \ \mu$eV for 30 nm separated dots\cite{dassarma2010exchange} and $j_v\ll1 \ \mu$eV\cite{culcer2010roughness}. The on-site repulsion in the right dot  $U_o^R=\bra{R_v R_{v(\bar{v})}}V_{ee}\ket{R_vR_{v(\bar{v})}}$, or charging energy, is of the order of few tens of meV\cite{lim2011spin,yang2012orbital}. The remaining two Coulomb integrals do not appear explicitly in $H_C$ since the direct Coulomb interaction $k=\bra{L_{z} R_{\pm z}}V_{ee}\ket{L_{z} R_{\pm z}}$ is an offset, while the Coulomb interaction enhancement terms $s=\bra{R_{\pm z} R_{z}}V_{ee}\ket{L_{\pm z} R_{z}}$ are part of the tunnel coupling $t$. It holds that $t=t_0+s$, where $t_0=\bra{R_{z}}H_0\ket{L_{z}}$. The last term in $H_{2e}$ is the tunnel Hamiltonian expressing the hopping of one electron between the two dots. The different terms of the Hamiltonian are:
\begin{align}
 H_0 &= \sum_{d=l,r}H_0^d
\;,
\label{eq:ham0}
\\ 
H_\epsilon &= -\epsilon \sum_{\substack{o=0,1\\ v=-,+\\ \sigma=\downarrow,\uparrow\\}}c_{\substack{R,o,\\ v,\sigma}}^\dagger c_{\substack{R,o,\\ v,\sigma}}
\;,
\label{eq:hamdet}
\\
H_C &= \sum_{o=0,1}U_o^R\sum_{\substack{o'=0,1\\ \sigma,\sigma'=\downarrow,\uparrow\\ v,v'=-,+}}n_{\substack{R,o,\\ v,\sigma}} n_{\substack{R,o',\\ v',\sigma'}}
\;,
\label{eq:hamC}
\\
H_t &= \sum_{\substack{v,v'\\ o,o'\\ \sigma}}t_{vv'}c_{\substack{R,o,\\ v,\sigma}}^\dagger c_{\substack{R,o',\\ v',\sigma}}\prod_{\substack{r=v_R,\sigma_R,o_R\\ R=S,V,O}}(-1)^{\delta_{r,ES}\delta_{R,0}}+H.c.
\;,
\label{eq:hamtun}
\end{align}
where $n$ is the number operator, $\sigma_R, v_R$ and $o_R$ are the spin, valley and orbital indexes of the right electron and $S,V,O$ are the spin, valley and orbital numbers of the two-electron state. The label $EV$ stands for the excited state of the quantum number expressed by $r$. The condition $S (V,O)=0$ means that the spin (valley or orbital) part of the 2-electron wavefunction is a spin singlet (valley or orbital) built from the single particle states. $t_{\pm\pm}$and $t_{\pm\mp}$ are the intravalley and intervalley tunnel couplings, respectively\cite{culcer2010roughness,petta2016dispersive_theory,petta2017dispersive_exp}. The first(second) coupling allows for tunneling between valley eigenstates of the same(different) form. We note that both terms prevent tunneling between states that have a different bulk valley index\cite{culcer2010multivalley}. As shown in Ref. \citenum{culcer2010roughness} these terms are complementary: one can increase only at the expenses of the other. It holds\cite{culcer2010roughness} that $t_{\pm\pm}=\frac{t}{2}[1+e^{-i\Delta\phi}]$ and $t_{\pm\mp}=\frac{t}{2}[1-e^{-i\Delta\phi}]$, where $\Delta\phi=\phi_L-\phi_R$ is the valley phase difference. The exact value of $\Delta\phi$ depends strongly on microscopic origins such as the interface roughness and the height difference between the dots. For instance, since $\phi=2k_0d$, where $d$ is the distance from the interface, even a single terrace step ($d=a_0/4$) leads to a quite large phase difference\cite{friesen2007valley} $\phi\approx0.84\pi$. In the case of a negligibly small height difference and a flat interface the valley mixing is the same and the valley eigenstates have the same valley composition. In practice, however, typical quantum dots have an orbital spacing on the order of 10 meV, corresponding to a dot size of around 10 nm, which is comparable to the correlation length range (few to hundreds of nm) reported for the Si/SiO$_2$ interface\cite{zimmerman2017valley}. As such, we expect different quantum dots to have different valley compositions.

\subsection{Two-electron energy levels}
Having considered 10 single-particle spin-valley states, the Hamiltonian $H_{2e}$ is expressed on a 26-state basis. These are the twenty-two lower orbital states and four $(0,2)$ states describing the same-valley double occupancy of the ancilla qubit. The latter one only occurs when accounting for the first orbital state. Therefore, the state $\ket{\downarrow,\downarrow}$ can evolve in a $(0,2)$ state via intravalley coupling. However, these same-valley double occupancy states contribute significantly to the eigenstates only at high detuning (i.e. $\epsilon\gtrsim U^R+E^R_o$) and we neglect therefore higher energy $(0,2)$ states. The basis states are tensor products\cite{Kondo2015Crippa,culcer2010multivalley} of the form $\ket{(\sigma_L,\sigma_R)}\ket{\psi_{V}}\ket{\psi_O}$ and $\ket{\chi_S}\ket{\psi_V}\ket{\psi_O}$ for the $(1,1)$ and $(0,2)$ charge configurations. Here $\ket{\chi_S}$ is the two-spin wavefunction while $\ket{\psi_V}$ and $\ket{\psi_O}$ are the symmetrized two-particle valley and orbital functions. For simplicity, from here on the orbital part is dropped, while we label the $(0,2)$ states as $S^{v,v'}_{(0,2)}$, $T^{v,v'}_{0,(0,2)}$ or $T^{v,v'}_{\pm,(0,2)}$.
\begin{figure*}
\includegraphics[width=\textwidth]{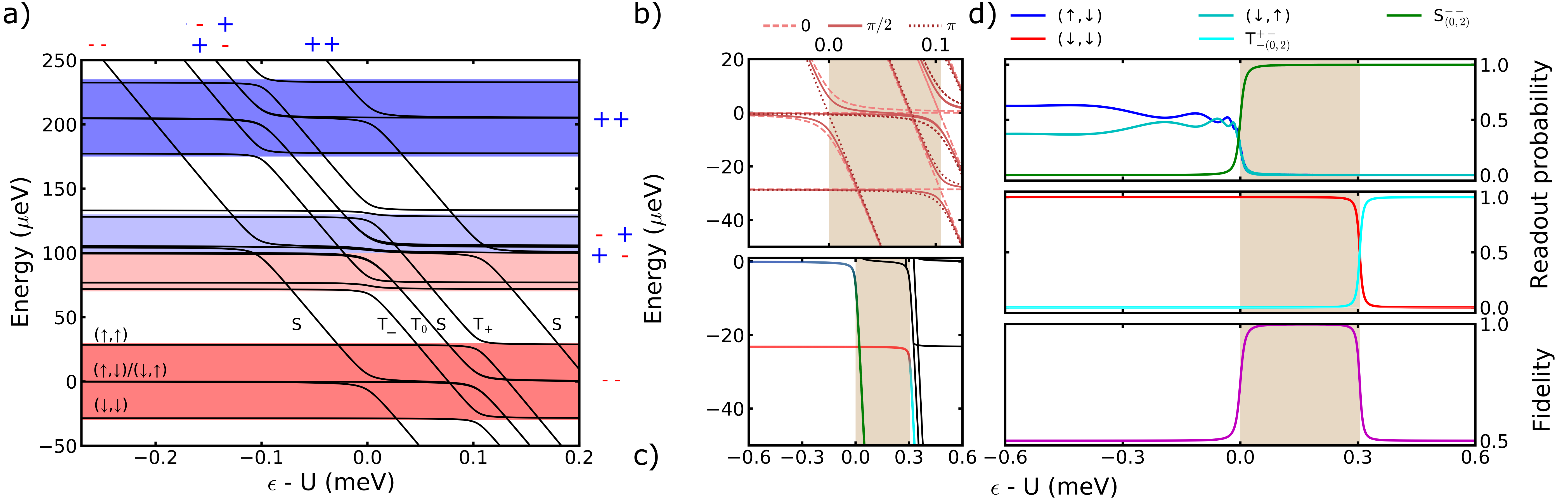}
\caption{a) Energy levels of a multivalley double quantum dot with $E_v>E_Z$. We distinguish three separate branches of energy states: the valley ground states (red), the valley excited states (blue), and the valley mixed states (light blue and red). The difference in valley splitting causes the splitting between the $(+,-)$ and $(-,+)$ sub-branches. The simulation parameters were: $U_o^R=30$ m$e$V, $t=1.5$ GHz, $\Delta \phi=\pi/2$, $E_v^{R(L)}=$ 105 (100) $\mu$eV, $E_o^R=$ 10 m$e$V and $E_Z^{R(L),-}$ = 28.65 (28.5) $\mu$eV. The two-spin states have a similar order for the three branches and are shown for the lowest branch only. b) Zoom-in at the intravalley anticrossing. Increasing the phase difference from 0 to $\pi$ changes the tunneling from pure intravalley to pure intervalley. The light brown rectangle highlights the high fidelity detuning range. c) Schematic of Pauli-spin blockade readout sequence. The states used in the readout protocol are shown with the same colors as in d). The double quantum dot is initialised either in $\ket{\uparrow,\downarrow}$ (in blue) or $\ket{\downarrow,\downarrow}$ (in red). The detuning is consequently changed linearly with an adiabatic pulse. Here $E_v^{R(L)}$ = 300 (305) $\mu$eV and $\delta E_Z^-$ = 20.7 neV. d) Results of time evolution simulations, using the same parameters as in c). Inside the high fidelity region $F$ is higher than 99.9\%. The oscillations at negative detuning in the top panel are the fingerprint of a singlet $(1,1)$. The pulse starts at the symmetry point and the pulse duration is set to achieve a high degree of adiabaticity (See Appendix for more details).}
\label{fig:levels}
\end{figure*}
\\In this work, we consider the case when $E_v^R\gtrsim E_v^L>E_Z^{R,L}$ and, as shown in Fig. \ref{fig:levels}a three branches emerge\cite{culcer2010multivalley}: in the lowest ($--$) and highest ($++$) branch the two electrons have the same valley number, while in the middle branch they are opposite ($\pm\mp$). The $(1,1)$ same-valley branches consist of four states each (i.e. $\ket{\downarrow,\downarrow}$, $\ket{\uparrow,\downarrow}$, $\ket{\downarrow,\uparrow}$, $\ket{\uparrow,\uparrow}$ in ascending energy order, since $E_Z^{R,v}>E_Z^{L,v}$), while in the $(0,2)$ configuration these same states include only the spin singlet state, because of the Pauli exclusion principle. Triplet states with same-valley can be formed only by involving higher orbital states, thus defining higher energy branches (not shown in Fig. \ref{fig:levels}a). The different-valley branch includes eight levels when in the $(1,1)$ and four states in the $(0,2)$ charge states. The difference in Zeeman energy sets the energy splitting between the antiparallel spin states in the three branches. The valley splitting energy sets the energy separation between the branches. A small difference in valley splitting energy splits the ($+-$) and ($-+$) states, as shown in Fig. \ref{fig:levels}a. The control of $\Delta\phi$ allows to select the nature of interdot tunneling, ranging from intra- to intervalley-only tunneling, as shown in Fig. \ref{fig:levels}b. In particular, for $\Delta\phi=0$ the $\ket{\downarrow,\downarrow}$ states are uncoupled from the $(0,2)$ charge states in the lowest orbital and the blocked region extends to the orbital spacing energy.\\ 

\subsection{Two-dot Pauli-Spin Blockade readout}
At negative detuning (i.e. $\epsilon\ll U^R$), the two lowest eigenstates can be approximated with the basis states $\ket{\downarrow,\downarrow}\ket{--}$ and $\ket{\uparrow,\downarrow}\ket{--}$. Differing only in the orientation of the target qubit spin, these states are hereafter used as initial states of the Pauli-spin blockade readout protocol and their valley label is dropped.\\ 
As shown in Fig. \ref{fig:levels}c, Pauli-spin blockade readout consists of a spin-to-charge conversion, which exploits the difference in charge configuration. At the beginning of the readout protocol, the ancilla qubit is in the ground state while the target qubit can be either spin up or spin down. The readout pulse detunes the double quantum dot beyond the intravalley anticrossing and inside the blocked region ($U^R<\epsilon_f < U^R+E^R_{v}-E_Z^{R,-}$), the brown region in Fig. \ref{fig:levels}c. As shown in Fig. \ref{fig:levels}d, if the two spins are initially antiparallel (blue level in Fig. \ref{fig:levels}c) the final state will be the singlet $S^{--}_{(0,2)}$ (green level); otherwise, the system will remain blocked in $\ket{\downarrow,\downarrow}$ (red level) until it relaxes via a spin flip. Experimentally, the final state can be probed either by charge sensing\cite{petta2005coherent,maune2012HRL} or by gate based dispersive rf-readout\cite{BetzNL}. However, these techniques require slightly different pulses. The former detects differences in the electric field, while the latter probes the level mixing via the quantum capacitance\cite{petersson2010DefinitionQC}. The highest fidelities are obtained far from or close to the intravalley anticrossing\cite{mizuta2017quantum}, respectively.\\
Here we consider linear adiabatic pulses conceived for charge sensing duration of 1 $\mu$s. (See the Appendix for details on pulse adiabaticity.) We note that shaped pulses could improve speed and performance (see Ref. \citenum{Roy_architecture} and therein references), although in arrays operated by shared control linear pulses could be required\cite{Roy_architecture}. The duration is chosen as a trade off between fast pulses and adiabaticity. The conversion fidelity $F$ is defined as the combined probability that $\ket{\uparrow,\downarrow}$ evolves to a $(0,2)$ state while $\ket{\downarrow,\downarrow}$ remains in a $(1,1)$ state: 
\begin{equation}
\begin{split}
F &= \frac{F_{\ket{\uparrow,\downarrow}\rightarrow(0,2)} + F_{\ket{\downarrow,\downarrow}\rightarrow(1,1)}}{2} 
\\
&= \frac{1}{2}\Bigg[\sum_{a\in(0,2)}|\braket{a|f}|^2+\sum_{b\in(1,1)}|\braket{b|f'}|^2\Bigg],
\label{eq:f}
\end{split}
\end{equation} 
where $f$ and $f'$ are the two final states calculated from the time evolution of the two lowest-lying eigenstates $\ket{\uparrow,\downarrow}$ and $\ket{\downarrow,\downarrow}$, respectively.\\
From Eq. \ref{eq:f} it can be seen that the ultimate limit to the readout fidelity is set by the final state composition, i.e. even a perfectly adiabatic pulse results in $F<1$ if the final state $f'$ has a non negligible contribution from $T_{-,(0,2)}^{+-}$ (see Fig. \ref{fig:levels}a). The effect of the phase difference is to change the eigenstate composition at a fixed detuning, potentially lowering the fidelity.\\
We recall that we have assumed negligible spin orbit coupling. Even though in bulk silicon the spin-orbit coupling is very weak, in quantum dots defined at the Si/SiO$_2$ interface the inversion asymmetry causes a finite spin-orbit coupling. The structural inversion asymmetry leads to a Rashba spin-orbit coupling, while the dominant Dresselhaus\cite{ruskov2017electron} arises from the interface inversion asymmetry\cite{golub2004spinorbit,nestoklon2006spinorbita,nestoklon2008spinorbitb}. The spin-orbit coupling strength depends, apart from the magnetic field orientation, on the vertical electric field, valley composition and the microscopic properties of the interface\cite{Carroll?}. In actual devices it can be non-negligible causing $g$-factor variability\cite{ferdous2017interface}, valley dependency\cite{ferdous2017anysotropy,veldhorst2015gfactor} and mixing between antiparallel and parallel spin states\cite{huang2017stepEDSR}. As a consequence, when including the spin-orbit Hamiltonian in $H_{2e}$ anticrossings between $S_{(0,2)}^{--}$ and the polarized triplets emerge\cite{fogartyPSB,Carroll?}. Further, such mixing would reduce $F$ even for adiabatic pulses. The shape of the pulse used for Pauli-spin blockade readout has to be modified accordingly, i.e. a two-speed linear pulse, to allow for a diabatic crossing of the $S-T^-$ anticrossing\cite{taylor2007adiab}. Therefore our assumption of negligible spin-orbit coupling ensures that our results demonstrating the impact of valley phase are not obscured by spin-orbit effects.

\section{Results}
\label{sec:results}
\begin{figure}
\includegraphics[width=\columnwidth]{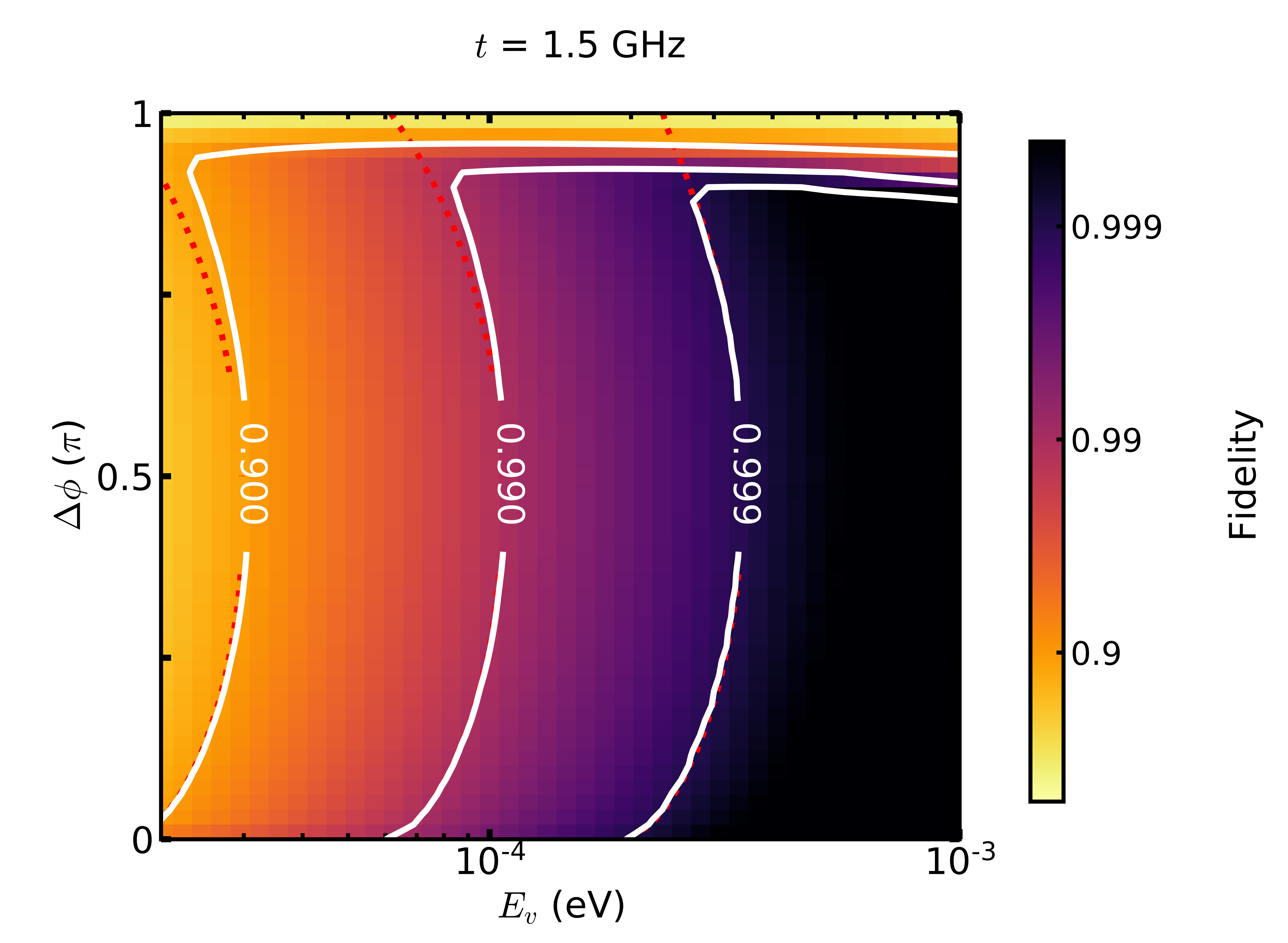}
\caption{Fidelity obtained by pulsing from $\epsilon = U^R -1$ meV to $U^R +E_v^R$ in 1 $\mu$s with $\Delta$t = 1 ps, $t$ = 1.5 GHz and $\delta E_Z^-=5$ MHz, for a range of experimentally achieved valley splitting energies (here $E_v^R=E_v^L$). Contour lines are shown in white. The decrease in $F$ for $\Delta\phi$ approaching $\pi$ (i.e. $t_{\pm\pm}\rightarrow0$) is caused by an increase in diabaticity, due to the constant pulse duration, absent in adiabatic evolution (red). For each point of the map, we have plotted the maximum achievable fidelity by taking the optimal detuning.}
\label{fig:phsp}
\end{figure}
\begin{figure}
\includegraphics[width=\columnwidth]{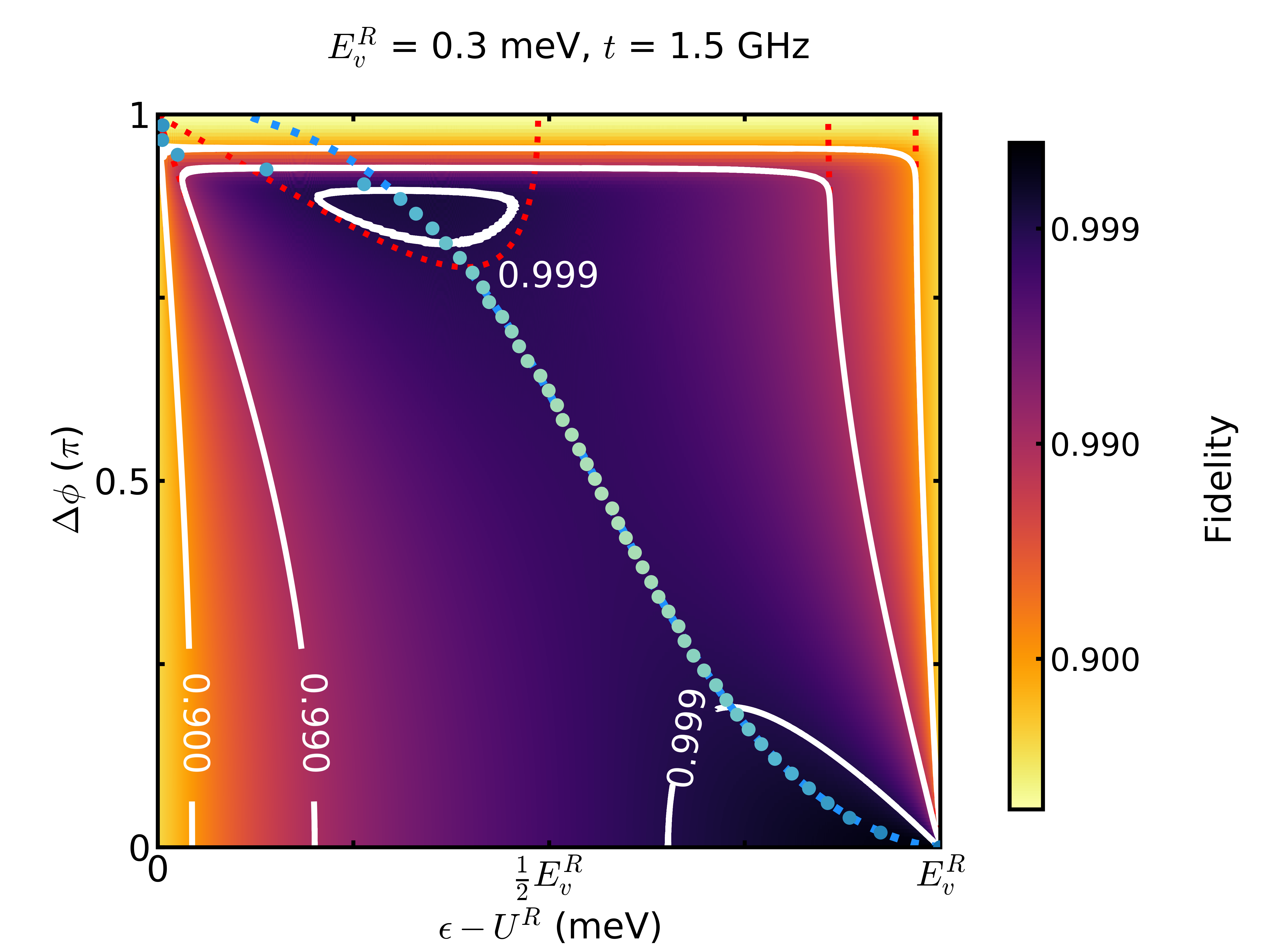}
\caption{Fidelity obtained by time evolution simulations (color map with white contour lines) and perfectly adiabatic pulses (red). The difference between the detuning position of maximum fidelity obtained from time evolutions simulations (dots) and adiabatic pulses (dashed line) is due to the finite speed of the pulse. As a consequence of the chosen parameters (e.g. $\delta E_Z^-=5$ MHz, $E_v^R = 300$ $\mu$eV, $t=1.5$ GHz and pulsing from $U^R -2E_v^R$ to $U^R +2E_v^R$) a gap where $F<99.9\%$ opens for intermediate $\Delta\phi$. The maximum fidelity has a non-monotonic dependence on $\Delta\phi$, visualized here by the color map of the dots; a minimum is observed at $\pi/2$ (light green), a local maximum at $\pi$ (blue), and the maximum at 0 (dark blue).}
\label{fig:PhPos}
\end{figure}

 
From previous considerations, it emerges that the larger $E_v^R$ the greater $F$. In general, $E_v^R$ can be tuned via a vertical electric field\cite{Morello_spinValley,veldhorst_addressable,saraiva2011EMAtheory}. In the device shown in Fig. \ref{fig:model}b, valley splitting can be controlled via the combined tuning of G3 and confinement gate C.\\
In Fig. \ref{fig:phsp} we show how the phase difference impacts on $F$ for different valley splittings (here $E_v^R=E_v^L$). Whenever $E_v^R>E_Z^{R,L}/2$, $F>80\%$ can be reached; in general we find a fidelity higher than 90$\%$ for $E_v\gtrsim 40t$. For a fixed valley splitting, the phase-dependence of $F$ is non monotonic, as visible for small splittings ($E_v<30 \mu$eV). At low $\Delta\phi$ the fidelity is high because the intervalley anticrossing is very narrow and the two final states have different charge configurations over a large detuning range. The minimum at $\Delta\phi\approx\frac{\pi}{2}$ arises from the opposite phase dependence of $t_{\pm\pm}$ and $t_{\pm\mp}$. Here a higher $E_v^R$ is needed to realize a large energy separation between the two anticrossings in order to reach the same fidelity (see 90\% contour line in Fig. \ref{fig:phsp}). For $\Delta\phi>\frac{\pi}{2}$ the fidelity increases with increasing phase, but now a high fidelity is achieved only for detunings close to the intravalley anticrossing (see Fig. \ref{fig:PhPos}). The decrease in $F$ at high $\Delta\phi$ is due to the increase in the pulse diabaticity. The conversion fidelity obtained from adiabatic pulses show that a fidelity higher than 90\% can be reached even for $\Delta\phi\approx\pi$, as highlighted by the dotted red lines in Fig. \ref{fig:phsp}, although it requires an impractical slow pulsing rate.\\
Properly tuning the readout position given a random phase difference is beneficial. The optimal readout point shifts with $\Delta\phi$ reflecting the state composition. As shown by the dots in Fig. \ref{fig:PhPos}, for a small phase difference it is convenient to readout close to the intervalley anticrossing, while for a large difference the pulse should end slightly beyond the intravalley anticrossing. Moreover, since $E_v^R/t\sim50$ a fidelity higher than 90\% is obtainable for very large detuning and $\Delta\phi$ ranges, except where the level mixing is strong (e.g. $\epsilon\sim0$ and $\Delta\phi\sim0$ or $\epsilon\sim E_v^R$ and $\Delta\phi\sim\pi$). On the contrary, there are two regions where $F>99.9\%$. At low $\Delta\phi$ the small $t_{\pm\mp}$ allows for reaching the detuning region where $\ket{\uparrow,\downarrow}$ is almost entirely converted to $S_{(0,2)}^{--}$ while the antiparallel spin state is still a $(1,1)$ state. Vice versa, for $\Delta\phi>\frac{\pi}{2}$ the increasing intervalley coupling is compensated by the smaller detuning needed for the $\ket{\uparrow,\downarrow}$ to evolve to $S_{(0,2)}^{--}$. Even the fidelity at the optimal point is a function of $\Delta\phi$, reflecting the state composition. In particular, the maximum fidelity is reached, as expected, for $\Delta\phi=0$ and $\epsilon=E_v^R$ while a minimum arises at $\Delta\phi=\pi/2$. Therefore the proper tuning of $\epsilon_f$ enables to reach 99\% fidelity threshold in a very large range of valley splittings and phase differences. 
\begin{figure}
\includegraphics[width=\columnwidth]{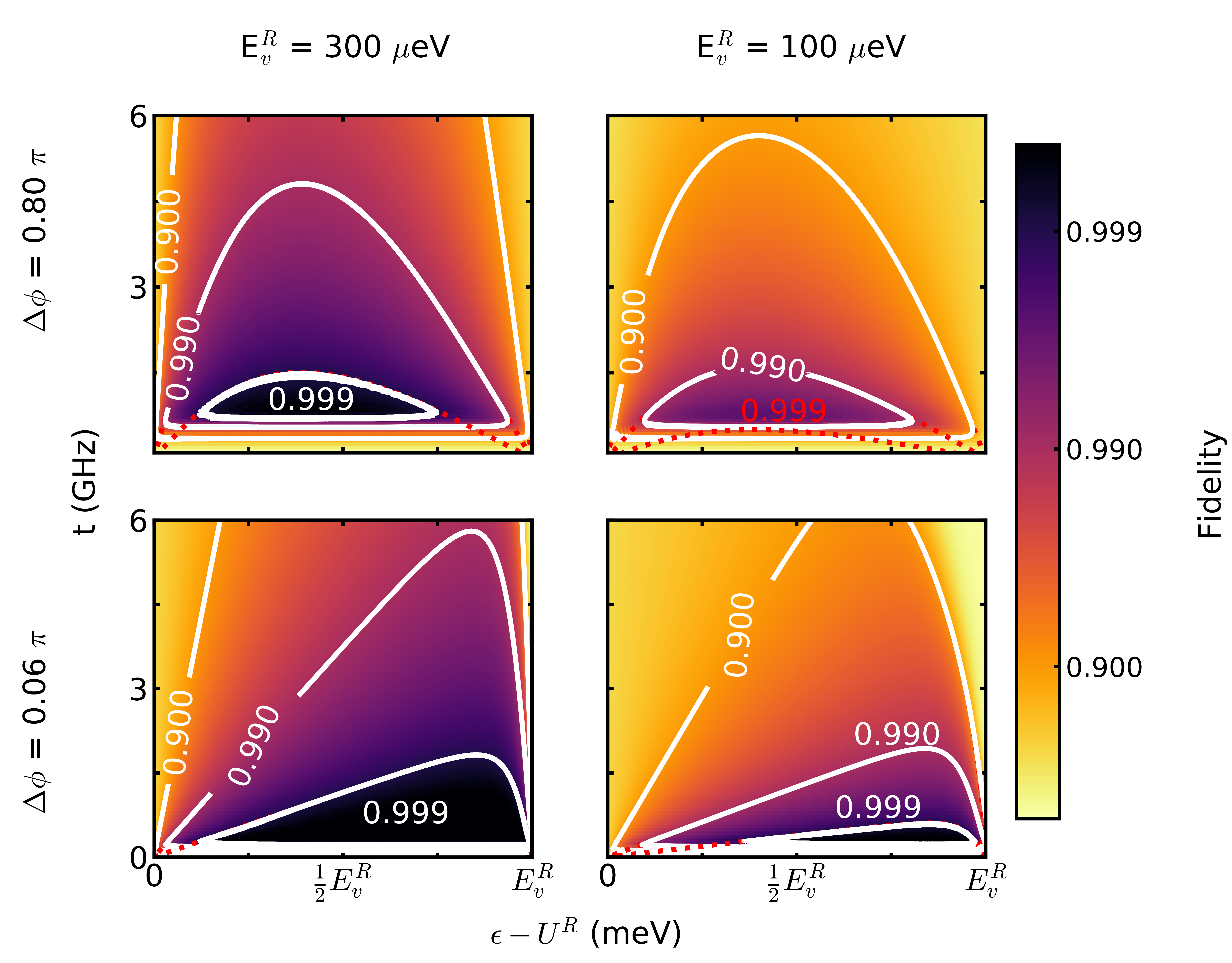}
\caption{Fidelity obtained by time evolution simulations (color map with white contour lines) and adiabatic pulses (red). A realistic value of 0.3 meV enables reaching $F>99\%$ for $t$ up to 4.5 GHz in a wide range of phase differences (here up to $\Delta\phi=0.8\pi$). The reduction in fidelity at low $t$ is caused by higher diabaticity of the pulse. In the top right panel it prevents reaching $F>99.9\%$. As in Fig. \ref{fig:PhPos} $\delta E_Z^-=5$ MHz and the pulse extremes are $U^R \pm2E_v^R$.}
\label{fig:Pht0}
\end{figure}
\\However, when aiming at $F>99\%$ or higher, only the control of the ancilla qubit valley splitting and/or $t$ enables overcoming the low fidelity region at intermediate $\Delta\phi$. The two 99.9\% regions merge for $E_v^R/t\sim54$, e.g. when $t=1.5$ GHz a valley splitting of at least 0.36 meV is required. Figure \ref{fig:Pht0} shows the fidelity as a function of phase and detuning for two valley splittings. For a valley splitting of 0.1 meV and considering perfect adiabatic pulses, a fidelity beyond 99.9\% can be reached for $t\lesssim500$ MHz and $0\leq\Delta\phi\leq0.7\pi$. When the valley splitting is slightly larger, i.e. 300 $\mu$eV, the same fidelity can be achieved for $t<1.5$ GHz. When the valley splitting is 700 $\mu$eV, a fidelity of 99\% can be reached when $t<5$ GHz and a fidelity of 99.9\% requires $t <3$ GHz.

\section{Triple dot readout protocol}
\label{sec:TQD}
\begin{figure}
\includegraphics[width=\columnwidth]{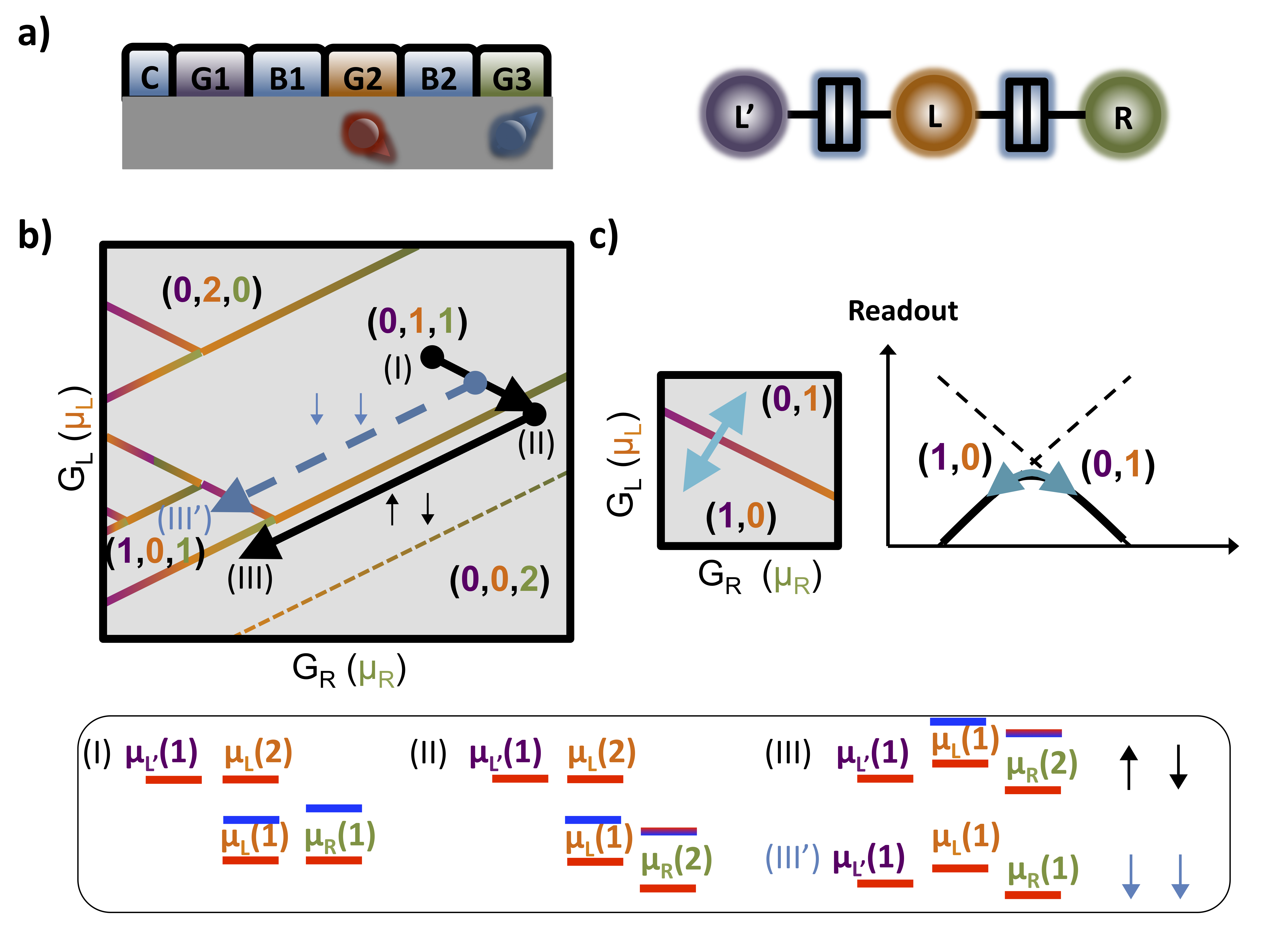}
\caption{Triple dot readout scheme. a) Schematic of a triple dot device with negligibly weak long-range coupling. b) The first step of the pulse (from (I) to (II)) is a standard double-quantum dot Pauli-spin blockade readout pulse. The spin state of the target qubit (orange) is converted to a spin-charge state via the right ancilla qubit (green). In the second step (from (II) to (III)), charge moves from the target to the left ancilla qubit (violet) only if the target qubit is originally in the spin-down state. The weak long range tunnel coupling locks the charges in the stable configurations $(0,0,2)$ and $(1,0,1)$, lifting the constrain set by spin relaxation time, while increasing the capacitive signal. Transitions between nearest neigbhoring dots are denoted by two-color interdot transition lines. The edges of the hysteresis regions are marked by three-color lines. The line color reflects the dots involved in the charge transitions. c) The readout is performed by oscillating the middle and left dot energy levels. Only the $(1,0,1)$ configuration leads to charge movement, thus allowing to measure the qubit state. Optionally, the tunnel coupling between the right dot and the middle one can be switched off. In this case the $(0,0,2)$ state will evolve to $(0,0)$ while $(1,0,1)$ to $(1,0)$.}
\label{fig:TQD}
\end{figure}

Experimental work on Pauli-spin blockade in silicon quantum dots show readout fidelity significantly lower than the conversion fidelities reported here\cite{fogartyPSB,harvey2017latching,tarucha2017readout}. This reduction is predominantly due to the small sensitivity of the charge sensor to variations in the electric dipole caused by a difference in the charge position. Therefore protocols based on a difference between the relaxation rates of singlet and triplet states\cite{studenikinEnhanced,fogartyPSB,harvey2017latching,tarucha2017readout} have been proposed. Usually a metastable state is exploited to achieve final states differing by one electron: the difference induced in the sensor signal by such variation is typically larger by a factor 1.4-4 and fidelities approaching the 99.9\% limit have been reported\cite{harvey2017latching,fogartyPSB}. The use of metastable states can lead, via a latching mechanism, to extremely long relaxation times.\\
Furthermore, when scaling up from few qubits to a large array ($N_{qubit}>1000$) it could be beneficial to split the spin-to-charge conversion from the actual readout process, allowing for delayed readout. Such separation can be achieved exploiting the latching mechanism reported for a double quantum dot\cite{dzurak2014hysteresis,harvey2017latching} asymmetrically coupled to a reservoir. Here we replace the reservoir with a third dot ($L'$) added to the left side of the double dot considered in the previous Sections, providing clear benefits in scalability. We assume a negligibly small long-range tunnel coupling $t^{RL'}$ between the two outer ancilla qubits (see Fig. \ref{fig:TQD}a, right panel). In the following we consider that the triple dot is loaded with two electrons at the beginning of the protocol and then the coupling to the electron reservoir is switched off. 
We assume that the triple dot is controlled by two ``virtual" gates G$_L$ and G$_R$, which are linear combinations of the B and G gates shown in Fig. \ref{fig:TQD}a. In particular, while G$_L$ shifts only $\mu_L$, G$_R$ shifts mainly $\mu_R$ and  $\mu_{L'}=0$, where $\mu_d$ is the chemical potential of the dot $d$. The condition $t^{RL'}=0$ alters the stability diagram similarly to the case of hysteretic double quantum dots\cite{dzurak2014hysteresis}. The interdot transition lines with a positive(negative) slope in Fig. \ref{fig:TQD}b satisfy the condition $\mu(1)_{L'(R)}=\mu(1)_L$. Further, here we assume that $\Delta\phi$, $t$, $\epsilon$ and $E_v^R$ have been optimized accordingly to the previous Sections.\\ 
The pulse protocol starts in a $(0,1,1)$ charge configuration. Then the system is detuned inside the $(0,0,2)$ spin-blocked window. Next, G$_L$ and G$_R$ are lowered together, raising the chemical potentials. The detuning direction is parallel to the $(0,1,1)\leftrightarrow(0,0,2)$ charge transition line so that $\mu(2)_R<\mu(1)_L$ and their relative offset is kept constant. The pulse ends deep in the $(0,0,2)$ region where $\mu(2)_R<\mu(1)_{L'},\mu(1)_L$, i.e. below the extension of the $(0,1,1)\leftrightarrow(1,0,1)$ transition line. The short-range coupling allows now for spin-to-charge conversion and charge shelving. If the separated spins were antiparallel, the $S_{(0,0,2)}$ will stay in the same charge configuration since when $\mu(1)_{L'}=\mu(1)_L$ no electrons are available for tunneling and the assumption $t^{RL'}=0$ blocks, at first order, the tunneling when $\mu(1)_{L'}<\mu(2)_R<\mu(1)_L$. On the contrary, parallel spins would still be in the $(0,1,1)$ charge configuration and when $\mu(1)_{L'}=\mu(1)_L$ an electron is transferred between $L$ and $L'$. As a consequence, $(0,\downarrow,\downarrow)$ and $(0,\uparrow,\downarrow)$ evolve to $(1,0,1)$ and $(0,0,2)$, locking the charge. Because of the the negligible long range tunnel coupling the spin flip relaxation time is now extend to a charge relaxation time, which happens via cotunneling. The next step of the protocol is the actual readout (Fig. \ref{fig:TQD}c). First the tunnel coupling $t^{LR}$ is completely switched off. The two possible final states of the $L'L$ double dot are $(0,0)$ and $(1,0)$, if at the beginning of the pulse the two spins were, respectively, antiparallel or parallel. Now rf gate-based dispersive readout can be used. The presence or absence of an electron in the $L'L$ double dot can be translated with high fidelity to the spin state of the target qubit. We note that  in the case of limited control of $t^{LR}$ this scheme can still be implemented, since the rf tone is applied such that the system oscillates between $(1,0,1)$ and $(0,1,1)$. Importantly, the possibility to doubly occupy the left ancilla qubit softens the experimentally demanding requirements of the triple donor scheme of Ref. \citenum{greentree2005readout}.
\section{Conclusions}
\label{sec:conclusions}
In summary, we have investigated the impact of (uncontrolled) valley phase difference on the Pauli-spin blockade spin-to-charge conversion fidelity. The damping effect of the phase can be mitigated by the control of the valley splitting of the ancilla qubit and additionally by tuning of the interdot bare tunnel coupling. In particular, we have shown that the control of the valley splitting energy together with the optimization of the readout position is sufficient to overcome randomness of the valley phase difference, even when the control of the tunnel coupling is limited and $t$ assumes realistic values. For $E_v^R>0.3$ meV a fidelity higher than 99.9\% can be reached for $t< 2$ GHz, as long as evolution is adiabatic with respect to the intravalley anticrossing. In addition, we have proposed a new protocol based on an isolated triple quantum dot to extend the Pauli-spin blockade readout measurement time by orders of magnitude, and significantly improving readout fidelity. 
\\Our results show that the randomness of the valley phase difference can potentially lower the readout fidelity. However, the experimentally demonstrated control of valley splitting and fine tuning of the detuning can overcome such variability. The extended relaxation time obtainable in a triple dot protocol makes Pauli spin blockade thereby an excellent method to be integrated in large-scale spin qubit systems.

\appendix*
\section{Adiabaticity threshold}
\label{sec:app}
\begin{figure}
\includegraphics[width=\columnwidth]{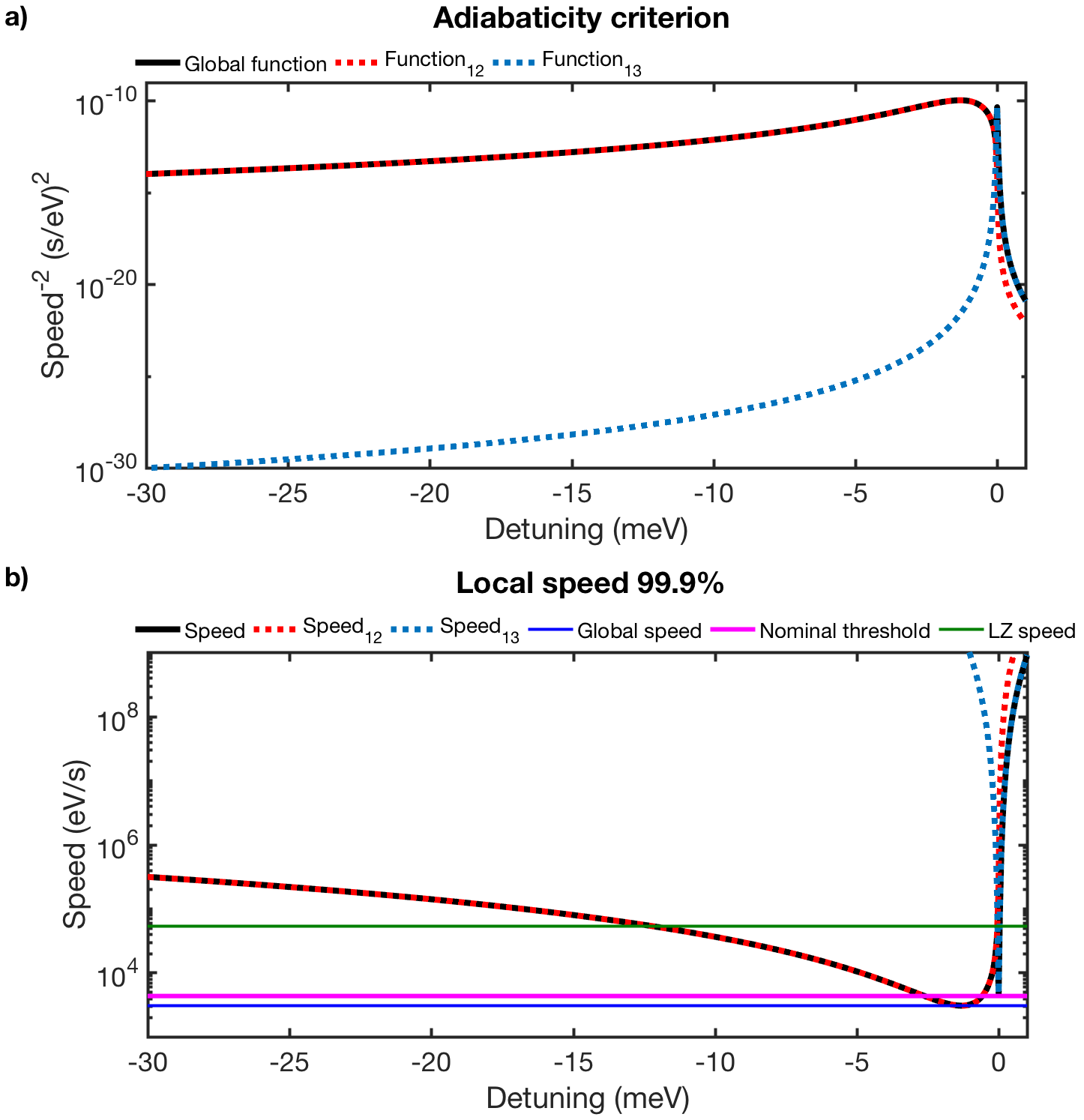}
\caption{a) The speed-normalized ``angular velocity" $\tilde{\alpha}_1$ of the ground state as a function of the detuning. At negative detuning it reduces to $\tilde{\alpha}_{12}$, while at positive to $\tilde{\alpha}_{13}$. Here $t$ = 1.5 GHz, $\delta E_Z=10$ MHz and $\Delta\phi = 0$. The peak at the anticrossing correspond to the contribution of a pure two-level system, while the broad peak is due to the fact that $\ket{\uparrow,\downarrow}$ and $\ket{\downarrow,\uparrow}$ are coupled to each other only via the singlet (0,2). b) The ``99.9\% adiabatic probability" local speed as a function of detuning. The lower and upper horizontal lines are the global speed obtained using Eq. \ref{eq:gen_cr} and the Landau-Zener formula, respectively. The middle one stems for the global 99.8\% global probability speed.}
\label{fig:appendix}
\end{figure}

In this Appendix we discuss the adiabaticity condition for a linear pulse. For two level systems a detailed theory has been developed and the Landau-Zener formula\cite{nori_landau,vitanov1996landau} $p=\mathrm{exp}\big(-4\pi^2t^2/hv\big)$ links the speed $v$ of a linear pulse to the probability $p$ of a diabatic transition between the eigenstates of the system. In the case of a multilevel system an analytical equation exists for the simple case of three-state ladder systems\cite{carroll1986ThLZ,vitanov2001review}, where two states are differently coupled to a third state and which successfully describes coherent adiabatic passage\cite{greentree_CTAP} or stimulated Raman adiabatic passage\cite{vitanov2017review}. 
\\Here we consider the three level system described by the Hamiltonian:
\[
H_{3L}= 
\begin{bmatrix} 
-\delta E_Z^-/2 & 0 & t_{--}\\
0 & \delta E_Z^-/2 & -t_{--}\\
t_{--}^* & -t_{--}^* & U_o^R-\epsilon
\end{bmatrix} 
\]
\\
written on the basis [$\ket{\uparrow,\downarrow}$, $\ket{\downarrow,\uparrow}$, $S_{(0,2)}$]. It approximates the 30-level system considered in the main close to the lowest valley branch intravalley anticrossing ($\epsilon\sim U^R$). Each of the three eigenstates $\Psi_{1,2,3}$ of $H_{3L}$ undergoes an adiabatic evolution when the criterion\cite{messiah_book}
\begin{equation}
\bigg|\frac{\alpha_i^\mathrm{max}}{\omega_i^\mathrm{min}}\bigg|^2\ll1
\label{eq:gen_cr}
\end{equation} 
is satisfied. Here $\omega_i^{min}$ is the minimum energy difference between the $i$-th eigenstate and the closest neighbour, while $\alpha_i^{max}$ can be seen as the maximum ``angular velocity''\cite{messiah_book} of the state $\Psi_i$ since it is defined as
\begin{equation}
|\alpha_i(t)|^2=\sum_{j\neq i}|\alpha_{ij}(t)|^2=\sum_{j\neq i}|\braket{\dot{\Psi}_i(t)|\Psi_j(t)}|^2.
\label{eq:alfa}
\end{equation} 
It has been shown (Ref. \citenum{messiah_book} for more details) that the total diabatic probability $p_i$ during the time evolution of the $i$-th eigenstate satisfies 
\begin{equation}
p_i\lesssim\mathrm{max}\bigg(\sum_{j\neq i}\bigg|\frac{\alpha_{ij}(t)}{\omega_{ij}(t)}\bigg|^2\bigg) <p_i^\mathrm{max}= \bigg|\frac{\alpha_i^\mathrm{max}}{\omega_i^\mathrm{min}}\bigg|^2.
\label{eq:pi}
\end{equation}
From Eq. \ref{eq:pi} the dependency of $p_i$ on the pulse speed can be obtained. Since for a linear pulse the speed $v=\dot{\epsilon}$ is constant we can rewrite $\dot{\Psi}_i(t)$ as $\dot{\Psi}_i(t)=\frac{\partial\Psi(t)}{\partial\epsilon}v$. An upper bound to the diabaticity probability is obtained by converting the inequality in Eq. \ref{eq:pi} to
\begin{equation}
p_i=v^2\mathrm{max}\bigg(\sum_{j\neq i}\bigg|\frac{\tilde{\alpha}_{ij}(t)}{\omega_{ij}(t)}\bigg|^2\bigg),
\label{eq:gen_cr_vel}
\end{equation} 
where $\tilde{\alpha}_{ij}(t)$ is the speed-normalized ``angular velocity". Equation \ref{eq:gen_cr_vel} can be used as a lower bound for the speed to obtain a defined $p_i$. 
\\In Fig. \ref{fig:appendix}a $\tilde{\alpha}_1$ is plotted, as well as the two contributions $\tilde{\alpha}_{12}$ and $\tilde{\alpha}_{13}$, as a function of the detuning. At negative detuning the dominant term is $\tilde{\alpha}_{12}$ meaning that the Zeeman energy difference sets the adiabaticity condition; while $\tilde{\alpha}_{13}$ better describes the system around zero detuning. For the particular case shown in Fig. \ref{fig:appendix} of $t_{--}=1.5$ GHz and $\delta E_Z=10$ MHz the peak at zero detuning is lower than the one related to the Zeeman energy difference. In such a scenario, it is possible to adiabatically pulse from $S_{(0,2)}$ to $\ket{\uparrow,\downarrow}$, defining the adiabaticity with respect to the tunnel coupling only, given a large detuning range and a small $p_1$. For higher tunnel coupling the zero-detuning peak becomes dominant and the two-level approximation becomes more accurate.
\\Eq. \ref{eq:gen_cr_vel} can be used to set the speed of a Pauli-spin blockade pulse in such a way that pulses with different $t$ satisfy the same adiabaticity condition. The function to be maximized in the right-hand side of Eq. \ref{eq:gen_cr_vel} can be viewed as a ``local" speed since it is a function of time and thus detuning. As shown in Fig. \ref{fig:appendix}b, it has the same trend as $\tilde{\alpha}_1$ and can be analogously split in two contributions. While the speed obtained from Eq. \ref{eq:gen_cr_vel} corresponds tot the global minimum of the ``local" speed, the global speed calculated from $p_i^\mathrm{max}$ is orders of magnitude smaller. Since $p_i^\mathrm{max}$ is an upper bound, using this definition will make the pulses much slower than what is required, and the use of $p_i$ allows for faster pulses. The fidelity of a $\ket{\uparrow,\downarrow}\rightarrow S_{(0,2)}$ pulse is limited by the adiabaticity of the charge transition, therefore the higher the adiabaticity the higher the fidelity. In general, the global speed derived using the Landau-Zener formula for $p=0.1\%$ would result in a fidelity approaching 99.9\%, while setting $p_i=0.2$ in Eq. \ref{eq:gen_cr_vel} or replacing $t$ with $t/2$ in the Landau-Zener formula allows for fidelity higher than 99.9\%.\\
In the time evolution shown in Fig. \ref{fig:levels}d a linear pulse from $\epsilon=0$ to $\epsilon=U^R+2E_v^R$, with the $p_i=0.2$ approximation, was used.

\acknowledgements
M.L.V.T. and M.V. gratefully acknowledge support from Intel. W.H. and A.S.D. acknowledge support from the Australian Research Council (CE11E0001017), the US Army Research Office (W911NF-13-1-0024) and the Commonwealth Bank of Australia.

\bibliography{\biblio}
\end{document}